\documentstyle[12pt]{article}

\large
\newcommand{\be}{\begin{eqnarray}}
\newcommand{\ee}{\end{eqnarray}}

\begin{document} 
\setlength{\baselineskip}{23pt}
\setlength{\baselineskip}{27pt}
\pagestyle{empty}  
\vfill                                                                          
\eject                                                                          
%vbox to  0.8in{} 
\vskip 0.5cm 
\renewcommand{\thefootnote}{\fnsymbol{footnote}}
\centerline{\bf\LARGE Hadronic Wave Functions} 
\centerline{\bf\LARGE in the Instanton Model}
\vskip 1cm
\centerline{\bf T.~Sch\"afer\footnote{supported in part by the 
Alexander von Humboldt foundation.} and E.V.~Shuryak}
\vskip 1.0cm                   
\centerline{\it Department of Physics}
\centerline{\it State University of New York at Stony Brook}
\centerline{\it Stony Brook, New York 11794, USA}
\vskip 1.0cm                                                                   
                                                                                
\centerline{\bf Abstract}
In this paper we wish to study hadronic wave functions
using an instanton model for the QCD vacuum. The wave functions
are defined in terms of gauge invariant Bethe Salpeter amplitudes
which we have determined numerically using a Monte Carlo simulation
of the instanton ensemble. We find that the pion and the proton, as 
well as the rho meson and the delta have very similar wavefunctions 
but observe a sizeable splitting between mesons or baryons with 
different spin. We compare our results with data obtained in 
lattice gauge simulations.  
\vfill                                                                          
\noindent                                                                       
\begin{flushleft}
SUNY-NTG-94/5
\end{flushleft}
\eject
\pagestyle{plain}
\renewcommand{\thefootnote}{\arabic{footnote}}
\setcounter{footnote}{0}
\setcounter{page}{1}

\vskip 1.5 cm
\centerline{\bf 1. Introduction  }
\vskip 0.5 cm
Hadronic correlation functions are an important tool in understanding
the complicated nonperturbative structure of the QCD vacuum
\cite{Shuryak_cor}. These correlation functions are defined as
the vacuum expectation value $\Pi(x)=<T(j_h(x)j_h(0))>$ of the time 
ordered product of two currents with the quantum numbers of a given
hadron $h$. At short distances, asymptotic freedom predicts 
that the correlation function is dominated by the propagation 
of free quarks. In this case, the correlator follows a simple 
power law behavior, $\Pi(x)\sim 1/x^6$ for mesons and $\Pi(x)\sim
1/x^9$ for baryons. Corrections to this behavior can be studied 
using the Operator Product Expansion.    

In general the correlation function can be expressed in terms of
the contributions of intermediate physical states. At large 
spacelike separations, the lowest resonance in a given channel
dominates and the corresponding mass and coupling constant can
be extracted from the exponential falloff of the correlator.
Roughly speaking, the coupling constant measures the probability
for all quarks inside the hadron to be at the same point. In order
to obtain more detailed information about the structure of the hadron 
the concept of hadronic wavefunctions was introduced in 
\cite{Chu_etal}. In the case of the pion this quantity is defined 
as the following gauge invariant Bethe-Salpeter amplitude
\be
 \label{pion_bs}
 \psi_\pi(y)&=&\int d^4x\, <0|\bar d(x)Pe^{i\int_x^{x+y}A(x')dx'}
 \gamma_5 u(x+y)|\pi>\, .
\ee
Here the two quarks are connected by a string of flux. One should note 
that in a relativistic theory the concept of a hadronic wavefunction is not  
uniquely defined. Other functions providing somewhat different
information include the Bethe-Salpeter
amplitudes in a fixed gauge or density-density correlation functions
(inside a given hadron). In practice the Bethe-Salpeter amplitude 
eq.~(\ref{pion_bs}) is extracted from the three point correlator
\be
\label{pion_cor}
 \Pi_\pi(x,y)&=&<0|T(\bar d(x)Pe^{i\int_x^{x+y}A(x')dx'}
 \gamma_5 u(x+y)\bar d(0)
 \gamma_5 u(0))|0> \sim \psi(y) e^{-m_\pi x}
\ee 
where $x$ has to be a large spacelike separation in order to 
ensure that the correlation function is dominated by the 
ground state and $y$ is the separation of the two quarks in 
the transverse direction. 

The Bethe-Salpeter amplitude eq.~(\ref{pion_bs}) has been measured in 
a number of lattice gauge simulations, both at zero and at finite
temperature \cite{Chu_etal,Negele,Bernard_etal,Schramm_Chu}. However, with the 
exception of one work in the context of the bag model \cite{Chu_etal_2} 
and the case of very large temperatures, where the wave function in the 
dimensionally reduced theory can be evaluated from an effective Schr\"odinger
equation \cite{Koch_etal,Hanson_etal}, there have been no efforts to
determine these wavefunctions in a theoretical model. 

We would like to fill this gap by applying the Random Instanton Liquid 
Model \cite{Shuryak_1982,Shu_Ver_1,Shu_Ver_2} to the problem of calculating 
hadronic wavefunctions. The model is based on the assumption that
the QCD vacuum is dominated by a particular type of gauge field 
configurations, small size instantons. The vacuum  is then characterized
by the density $n_0=1\;{\rm fm}^4$ and the average size $\bar\rho=1/3\;
{\rm fm}$ of the instantons. Essentially, these numbers were chosen 
to reproduce two global properties of the QCD vacuum, the value of the
quark and gluon condensates. Recently, we have determined a large
number of hadronic two point correlation functions using this model
\cite{Shu_Ver_2,Schaefer_etal}. The model not only successfully describes 
the masses and coupling constants of the lowest resonances in most channels, 
but also reproduces the results of a recent lattice calculation of two 
point correlation functions \cite{Negele_etal}.

In the present work we want to extend this comparison to hadronic
Bethe-Salpeter amplitudes. First of all, this will enable us to check whether 
the correlators calculated in \cite{Shu_Ver_2,Schaefer_etal} really 
correspond to the propagation of hadronic {\it bound} states. In addition
to that, one can study whether the size and shape of the corresponding
hadronic wave functions is similar to what was found in lattice 
simulations. In general, these are very non trivial questions. The 
Instanton Model takes into account only one very specific 
non perturbative effect among the many that determine the 
structure of QCD. Whether instanton induced effects really
dominate the physics of a particular observable has to be checked
in each individual case.

  Another interesting question concerns the  {\it  differences} 
between the wavefunctions of hadronic states with different quantum 
numbers. In a potential model approach, these differences are due 
to the gluomagnetic forces that give rise to a spin-spin interaction 
between quarks. These forces are absent in our calculations. However,
there also exists a spin-dependent instanton induced interaction 
between quarks, which qualitatively describes the mass splitting between 
different states. The question is whether it can also account for 
the differences in the size of the corresponding bound states. 

Finally we would like to address a few questions related to the 
interpretation of lattice results on hadronic Bethe-Salpeter
amplitudes. In particular we want to study the questions  
whether the observation of localized Bethe Salpeter
amplitude uniquely reflects  on the existence of a hadronic
bound state and whether measurements of correlation functions
at rather short distances on the order of one fermi (or even less)
are sufficient to deduce properties of the ground state in a given
channel.

\vskip 1.5 cm
\centerline{\bf 2. Wave Functions in the Random Instanton Model }
\vskip 0.5 cm
In the following we will study Bethe-Salpeter amplitudes for 
the pion, the rho meson, the nucleon and the delta resonance.
For this purpose we have considered correlation functions
involving the following currents
\be
  j^\pi (x) &=& \bar d(x)\gamma_5 u(x), \\
  j^\rho_\mu (x) &=& \bar d(x)\gamma_\mu u(x), \\
  j^p_\alpha (x) &=& \epsilon^{abc} (u^a(x)C\gamma_5 d^b(x)) 
  u^c_\alpha (x) ,\\
  j^\Delta_{\alpha,\mu}(x) &=& \epsilon^{abc} \left(
  2(u^a(x)C\gamma_\mu d^b(x))u^c_\alpha (x) 
  +(u^a(x)C\gamma_\mu u^b(x))d^c_\alpha (x) \right).
\ee
Here $q^a_\alpha$ denotes a $u$ or $d$-quark spinor with color index
$a$ and spinor index $\alpha$ and $C$ is the charge conjugation  
matrix. The nucleon current given above is a linear combination  
of the standard Ioffe currents \cite{Schaefer_etal}. Our choice
here is motivated by the fact that this current was used in the 
lattice simulations reported in \cite{Chu_etal}. We have done 
calculations using all six nucleon correlation functions introduced 
in \cite{Schaefer_etal} and will comment on the importance of the 
choice of current below. 

We consider the case of exact isospin symmetry so we have simply
chosen the flavor content of the mesonic currents to minimize 
the number of contractions contributing to the correlation function.
In the case of the nucleon we have used a proton current and defined
the corresponding wave function by taking the two $u$ quarks at the
same point and measuring the correlator as a function of the 
separation of the $d$ quark. Similarly, for the delta resonance
we have taken the $\Delta^+$ state and again measured the Bethe
Salpeter amplitude as a function of the distance between the two 
$u$ quarks and the $d$ quark. 

The Bethe-Salpeter amplitude defined in eq.~(\ref{pion_bs}) involves
a diagonal correlation function of two pseudoscalar currents. The same 
information, however, can also be obtained from non diagonal
correlation functions. In the case of the pion, we use  
the pseudoscalar-axialvector correlation function
\be
\label{ps_av}
 \Pi_{\pi-a_1,\mu}(x,y)&=&
 <0|T(\bar d(x)Pe^{i\int_x^{x+y}A(x')dx'}\gamma_5 u(x+y)\bar d(0)
 \gamma_\mu\gamma_5 u(0))|0>, 
\ee
while in the case of the rho meson, we have also determined the
Bethe-Salpeter amplitude from the off diagonal vector-tensor 
correlation function
\be
\label{v_t}
\Pi_{\rho-T,\mu}(x,y)&=&
 <0|T(\bar d(x)Pe^{i\int_x^{x+y}A(x')dx'}\gamma_\nu u(x+y)\bar d(0)
 \sigma_{\mu\nu} u(0))|0>. 
\ee
Similarly, for the nucleon one can use non diagonal correlation
functions between the two different Ioffe currents \cite{Schaefer_etal}.
The advantage of using non diagonal correlators lies in the fact 
that these correlation functions receive no contribution from free
quark propagation. For this reason the dominance of the ground state
over excited states at large but finite separation is expected to
be somewhat better for non diagonal correlators. In any case, 
comparing the Bethe Salpeter amplitudes determined from diagonal
and non diagonal correlators provides a nice check on whether the
measured wavefunctions really do reflect on the groundstate in the
given channel.

In addition to the channels considered here it would also be interesting
to study the wave functions of other mesons. Of particular interest
in comparison with the rho meson is the $a_1$ meson. This channel, 
however, is difficult to analyze since the $a_1$ correlation function
at large distances is dominated by the pion \cite{Shuryak_cor} and
the off diagonal $a_1$-tensor correlation function turns out to be
very small. Furthermore, it would be nice to demonstrate that one
does not observe localized Bethe Salpeter amplitudes in channels
in which (at least inside the model) no bound states are expected
to form. In the Instanton Model, one such channel would be the 
scalar isovector ($\delta$-meson) channel. Again, the 
analysis turns out to be very difficult since the fact that the
interaction is repulsive implies that the measured correlation
function at large distances is very small so that the wave function
cannot be reliably determined.

  The instanton ensemble in QCD can be generally described
by a partition function of the type
\be
  Z = \int \Pi_i [d\Omega_i \exp(-S_i)]\,\exp(-S_{int})\,
    \Pi_{f=1,N_f} {\rm det} (i\hat D+im_f)
\ee
where $[d\Omega_i]$ denotes the integration over the collective 
coordinates (orientation, size, position) associated with the 
instantons, $S_i=8\pi^2/g^2(\rho)$  is the action of an individual 
instanton and $S_{int}$ is the instanton-instanton interaction.
Since the evaluation of this partition function still constitutes 
a formidable problem we consider the even simpler Random Instanton
Model. In this model we assume that except for the size, which
we keep fixed, the distribution of the collective coordinates 
is completely random. In particular,
adopting this assumption we neglect the effects of the
fermion determinant ${\rm det}(i\hat D+im_f)$, which
is similar to ``quenched" lattice simulations we are going to compare
with.

The Bethe-Salpeter amplitudes are then determined from the quark
propagator in a given configuration. The pion correlation function
eq.~(\ref{pion_cor}), for example, is given by 
\be
 \Pi_\pi (x,y)&=&<{\rm Tr}S^{ab}(x+y)
 (Pe^{i\int_x^{x+y}A(x')dx'})^{bc} S^{ca}(-x)>
\ee
where the trace is performed with respect to the Dirac indices. The 
calculation of the propagator is explained in detail in \cite{Shu_Ver_2}.
Here we only outline the general strategy. The quark propagator in  
the field of a single instanton can be written as the sum of a zero
mode contribution and the propagator due to non zero modes. Both
pieces are known analytically. In the many instanton configuration 
we treat the zero modes exactly by numerically inverting the Dirac 
operator in the space spanned by the zero modes of the individual
instantons. The non zero modes, on the other hand, are taken into account 
only approximately, by including the first term in a multiple
scattering expansion. The resulting quark propagator is given by
\be
 S(x,y) = \sum_{IJ} <x|\phi_I>
 <\phi_I|\frac{1}{i\gamma_\mu D_\mu + im}|\phi_J>
 <\phi_J|y> + S_{NZM}(x,y), 
\ee
where $\phi_I(x)$ is the zero mode associated with the instanton $I$
and 
\be 
\label{snzm}
S_{NZM}(x,y) = S_0(x,y) + \sum_I (S^I_{NZM}(x,y)-S_0(x,y)) 
\ee
is the propagator due to non zero modes. Here, $S_0$ denotes the free
propagator and $S^I_{NZM}$ is the non zero mode propagator in the 
field of an individual instanton. The non zero mode propagator also
has to be corrected for the effects of current quark masses 
\cite{Shu_Ver_2}. In this paper we take the light quark mass to
be $m_{u,d}=10$ MeV and the corresponding effects are small.
The Schwinger factor $P\exp(i\int A_\mu dx_\mu)$ is calculated 
numerically for any given gauge field configuration by expanding 
the pathordered exponential as an infinite product
\be
\label{pexp}
 P\exp(i\int A_\mu dx_\mu) = \Pi_{i} (1+iA_\mu(x_i)d{x_i}_\mu)\,.
\ee
In practice this expression is approximated by a finite product
where the stepsize $dx_i$ is determined by the magnitude of 
the local gauge potential. 

In general, the inclusion of the Schwinger factor eq.~(\ref{pexp}) 
is expected to give an important contribution to the measured wave 
functions, since it corresponds to an additional string type potential. 
The instanton liquid, however, does not produce confinement and 
no string potential is expected to appear. At large separations, the 
Schwinger factor decreases exponentially $P\exp(i\int_0^r A\,dx)
\sim \exp(-mr)$ with $m$ being the mass renormalization of a heavy 
quark in the instanton background \cite{Schaefer_etal}.
Using the `sum ansatz' for the gauge field of the multi-instanton 
configuration we find $m=65$ MeV, in agreement with estimates 
made in \cite{Diakonov_etal_1989}. Employing the more
realistic ratio ansatz an even smaller screening mass $m=20$ MeV
is found. In any case, the insertion of the path ordered exponential will
only lead to minor effects on the wave functions we are going to discuss.
            
\vskip 1.cm
\centerline{\bf 3. Results and comparison with Lattice Gauge Simulations} 
\vskip 0.5cm
Before we come to the results obtained, let us discuss
the Bethe Salpeter wave functions obtained by using simplified 
models for the propagator. Since in all numerical simulations  to
date the separation between the sources is rather modest (at most 
on the order of 1 fm) it is important to understand what the size
of non asymptotic effects in the measured wave functions are.

In the simplest case we consider a free fermion propagator 
\be
  S_0(x)= \frac{i}{2\pi^2}\frac{\gamma\cdot x}{x^4}
\ee
and take the path ordered exponential to be unity. Here all gamma matrices
and coordinates are taken to be euclidean. Inserting this propagator
in the pion correlator eq.({\ref{pion_cor}) we get\footnote{Since we
are discussing short range effects, we are using a symmetric 
configuration in which the quarks are located at transverse 
separations $-y/2,+y/2$. For asymptotic wave functions, this 
difference should not matter.}
\be 
  \Pi_\pi(x,y)&=&\frac{N_c}{\pi^4}\frac{x^2-\frac{y^2}{4}}
  {(x^2+\frac{y^2}{4})^4}\simeq 
  \frac{N_c}{\pi^4x^6}\cdot (1-\frac{5}{4}\frac{y^2}{x^2}+\ldots).
\ee
At distances $\tau=\sqrt{x^2}$ on the order of one fermi, this  
corresponds to a 'wave function' with a mean square radius\footnote{ 
The mean square radius is defined by $\langle r^2\rangle=
\int d^3r r^2 |\psi(r)|^2$ where the Bethe Salpeter wave function
is normalized to one.}
$\sqrt{\langle r^2 \rangle}=0.89$ fm, quite 
different from the asymptotic result.  This effect 
becomes even more pronounced if one considers the propagation
of a massive constituent quark 
\be 
 S(x) = i\frac{\gamma\cdot x}{\tau}D'(m_q,\tau)
        + m_q D(m_q,\tau)
\ee
where $D(m,\tau)=m/(4\pi^2\tau) K_1(m\tau)$ is the coordinate space
propagator of a massive scalar particle. Taking the quark mass to be 300
MeV, we find a wave function with a size of $\sqrt{\langle r^2 \rangle} 
=0.79$ fm at a distance of $\tau=1.0$ fm. Even at a distance of 1.5 fm, 
the wave function is still fairly well localized with a radius of 
$\sqrt{\langle r^2\rangle}=1.1$ fm.
Note however, that so far the appearance 
of a localized wave function is a purely geometric effect and 
has nothing to do with the existence of bound states.

One should note that the fact that the Bethe Salpeter wavefunctions have 
not reached their asymptotic shape can not be detected from an analysis 
of the spacelike screening mass. The screening masses in this simple model
are $m_{scr}=2m_q$ for mesons and $m_{scr}=3m_q$ for baryons. There are only
small power corrections to this behavior that are due the fact that the 
quarks can have some transverse momentum.
We would also like to emphasize that considering the effects
of finite quark masses on the Bethe Salpeter wave functions is very
important at finite temperature where quarks develop thermal screening
masses $m_{scr}=\pi T$. Around $T_c$ this mass is on the order of 
500 $MeV$. If the wave functions are determined at distances $x\sim 1/T$,
which is typical for current lattice simulations, masses will have
an important effect on the measured Bethe Salpeter wave functions.

After these remarks concerning non asymptotic effects we would like 
to study how instantons lead to the formation of bound states that 
manifest themselves in localized asymptotic Bethe Salpeter amplitudes. 
A simple one-instanton approximation that has been used \cite{Shuryak_A4}
to analyze the qualitative effects of instantons on the {\it short-distance} 
behaviour of correlation functions is based on the following expression 
for the propagator in a dilute instanton gas  
\be
\label{zm}
 S(x)&=& S_0(x)+n_0\int d^4z\, 
    \frac{\phi_{0,z}(x)\phi^\dagger_{0,z}(0)}{-im^*}\,.
\ee
Here $n_0$ is the density of instantons, $\phi_{o,z}(x)$ is a zero mode
localized at $z$ and $m^*=2\pi\rho_c\sqrt{n_0/6}$ is the effective quark
mass in the instanton gas. Recall, that
such propagator eq.~(\ref{zm}) leads to an 
attractive pion correlation function and explains the failure of
the operator product expansion in this channel. At distances greater
than $\tau=0.5$ fm, however, the full correlation function is strongly 
underestimated. In order to produce a real pion bound state one has
to consider the corresponding four quark vertex and iterate this 
interaction in an RPA type approximation \cite{Diakonov_Petrov}.

The zero modes are localized around the centers of the instantons
with a characteristic size $\rho$. Evaluating the pion wave function 
eq.~(\ref{pion_bs}) using the propagator eq.~(\ref{zm}) we find
\be
  \psi_\pi(y) = 1-y^2/(2\rho)^2 + \ldots
\ee
for small transverse separations $y$. The corresponding mean square 
radius is $\sqrt{\langle r^2\rangle}=1.41$ fm at $\tau=1$ fm, but
does not converge as $\tau\to\infty$. This means that individual
instantons produce some interaction at short distances but the  
corresponding wave function has a long range tail and no true
bound state exists. 

A similar calculation can also be done for the nucleon. 
Using the proton current defined in eq.~(5) and the model for
the propagator introduced above, the nucleon correlation function
simply factorizes into the product of a free quark propagator
and a scalar diquark correlation function. The scalar diquark
correlator is very similar to the pion one, in particular it
is also attractive and the corresponding wave function as 
calculated from the propagator eq.~(\ref{zm}) is the same.
The isospin structure of the proton current is such that the
$d$-quark which is used to define the Bethe Salpeter wave function
is always in the scalar diquark. We then find that in this
approximation the nucleon Bethe Salpeter wave function is identical 
to the one of the pion: $\psi_N(y)=\psi_\pi(y)$. This 
fact is essentially a consequence of the diquark structure of
the nucleon current.

We have seen that such {\it single-instanton} approximation leads 
to a short range attraction which affects the wave functions, although 
no true bound state in the pion and nucleon channels appears. This 
is even more so for the rho meson and delta channels, in 
which individual instantons do not affect the correlator at all. 
In order to study the question whether an {\it ensemble} of 
instantons leads to the formation of hadronic bound states
we have done numerical simulations in the Random Instanton Model.
For these simulations we have studied an ensemble of 256  
instantons in a box $6.7\times 3.4^3\,{\rm fm}^4$. In order to 
minimize finite size effects while still being able to study large  
separations, the correlation functions are measured along the long axis 
(the 'time' direction) of the box. For the measurement of wavefunctions,
the available statistics limits us to the separations smaller than
1.5 fm.

The results of our calculations are shown in figures 1.-4. In 
fig.~1.~we show the measured pion Bethe Salpeter wave function
as a function of the longitudinal separation $\tau$. The wave
functions changes quite significantly between $\tau=0.5$ fm 
and $\tau=1$ fm, but remains roughly unchanged for $\tau>1.0$
fm. We have extracted the  corresponding mean square radii 
by assuming that the wave functions are exponential for
transverse separations $y>1.0$ fm. The pion rms radius measured
this way is $\sqrt{\langle r_\pi^2\rangle}=0.61$ fm at $\tau=1.25$
fm, significantly smaller that what we found for the dilute case 
considered above. 

\begin{table}
\caption{Rms radii extracted for the Bethe Salpeter wave functions
measured at a longitudinal separation $\tau=1.25$ fm. The various 
correlation functions are described in the text.}
\begin{center}
\begin{tabular}{|c||c|c|c|c|}\hline
           & $\pi$(P-P)  & $\pi$(P-AV) & $\rho$(V-V) & $\rho$(V-T) \\ \hline
$\sqrt{\langle r^2\rangle}\;[{\rm fm}]$ 
           &   0.61      &    0.56     &   0.73      &   0.70 \\ \hline\hline
    & $N(\eta_1-\eta_1)$ & $N(\eta_1-\eta_2)$ & $\Delta$  &   \\     \hline
$\sqrt{\langle r^2\rangle}\;[{\rm fm}]$
           &   0.63      &    0.61     &   0.70      &        \\ \hline
\end{tabular}
\end{center}
\end{table}

The fact that the Bethe Salpeter wave function appears to be 
dominated by the pion is also illustrated by a comparison of
the wave functions determined from the diagonal and off diagonal 
(pseudoscalar-axialvector) correlation functions (see fig.~2. and
table 1.). The two wave functions are very similar with the off 
diagonal one being somewhat smaller. As explained in section 2 this 
might be due to the fact that there is still some contribution
from the continuum left in the diagonal correlator while it is
absent in the off diagonal one.

In fig.~2.~we also show a similar comparison for the rho meson
wave function determined from the vector and vector-tensor
correlators as well as the nucleon wave function measured from the
diagonal and the off diagonal correlator of the first 
and second Ioffe current. In both cases the agreement is 
excellent and we conclude that the wave functions are likely
to be resonance dominated. This can also be seen from the 
fact that even in the rho meson channel, in which there
is no first order instanton induced interaction, the wave
function is smaller than what would be expected from the 
propagation of massive constituent quarks. 

A comparison of the wave functions obtained in different channels
is shown in fig.~3. We observe that the pion and the proton 
as well as the rho meson and the delta resonance have very similar
wave functions. The pion and the proton, however, are significantly 
smaller than the rho meson and the delta resonance. We have already
argued that the scalar diquark content of the nucleon and the fact
that instantons produce substantial attraction in the scalar diquark
channel suggest that the nucleon and pion wave functions are very 
similar. Analogously, one can think of the delta as consisting of 
a quark and a vector diquark. In this case, however, the above 
argument does not really apply since the $d$ quark is not 
necessarily sitting in the diquark and there is no interaction
that would enhance the importance of the quark-diquark part of
the correlation function.

In fig.~4.~we compare our results for the pion and the proton with
the lattice result reported in \cite{Chu_etal,Negele}. These authors
have determined the Bethe Salpeter wave functions for the pion, the  
rho meson and the nucleon in a quenched  lattice simulation.
In addition to that, they have also measured wave functions
after 'cooling' their configurations. This procedure relaxes any
given gauge field configuration to the closest classical solution
\cite{Chu_etal_3}. It removes essentially all gluon 
fluctuations except for those stabilized by topology. 'Cooling' 
therefore displays the contribution of instantons to the measured 
wave functions. Comparing the different curves shown in fig.~4.~we
observe that the full lattice wave functions are more compact than
the instanton results while the cooled wave functions are even  
larger. Qualitatively it is clear why the wave functions in the 
instanton model extend farther out: the instanton model misses at 
least two  effects, the Coulomb force at short distances and the string 
potential at large distances. Both help to localize the wave functions,
although none of them seems to be of crucial importance.

This can be seen in more detail by considering the shape of the wave functions.
At short distances  our wave functions are quadratic in the separation, while 
the lattice results appear to be linear. This effect should be due to the
perturbative Coulomb and spin-spin forces, since the linear behavior 
disappears when the wave functions are cooled. This can be checked further 
by applying only a few cooling sweeps (compared to the 25-50 sweeps used to 
extract the instanton liquid): the perturbative ``quantum noise"  disappears 
already at this level.

  Finally, although the shapes are similar, the absolute size of the wave
functions extracted from our model and from cooled lattice configurations
do not agree. A possible reason for this behavior is the fact that there 
are some ambiguities related to the redefinition of the lattice scale 
after cooling. Even before cooling, the lattice scale depends on the
particular observable that is used to fix the units \cite{Chu_etal}.

\vskip 1.5 cm
{\bf \noindent 4. Conclusions \hfil}
\vskip 0.5cm
We have measured Bethe Salpeter wave functions for the pion, the proton,
the rho meson and the delta resonance in the Random Instanton Liquid Model.

In all the channels considered here there is evidence that this 
Model leads to the formation of localized wave functions. This is based
on the fact that the wave functions become stable as the longitudinal
separation is increased and the observation that the wave functions
measured using different sources are very similar to each other.

Furthermore, the size of these bound states are quite similar to
those observed experimentally and measured on the lattice. In fact,
the model reproduces most of the qualitative features
of the wave functions that are observed in lattice simulations. In 
particular, we find that the wave functions of the pion and the proton  
are very similar whereas the rho meson has a significantly larger size.

  The shape of the wave functions measured on the lattice are somewhat 
different, though. This is an important observation since the model does 
not include some of the physics both at very short and at very large 
distances, namely the perturbative and the confining forces. The 
size of the observed deviations is therefore a measure for the importance
of these effects in the formation of hadronic bound states. After cooling
the qualitative behavior of the lattice results agrees with what is
observed in the Instanton Model. However, the mean square radii determined 
in the Instanton Model are somewhat smaller
as compared to the cooled lattice reported in \cite{Negele}. 

In conclusion we have shown that the Random Instanton Liquid
Model can account for the qualitative behavior of hadronic Bethe 
Salpeter wave functions. It will therefore be of great interest to extend  
the calculations presented in this work to include the effects of
finite temperature.

\vskip 1.5 cm
{\bf \noindent 5. Acknowledgements \hfil}
\vglue 0.4cm
 The reported work was partially supported by the US DOE grant
DE-FG-88ER40388. We acknowledge the NERSC at Lawrence Livermore where
most of the computations presented in this paper were performed.
We would like to thank Jac Verbaarschot for his collaboration on the
Instanton Model, and very helpful discussions with John Negele on lattice 
results.

\newpage
\begin{figure}
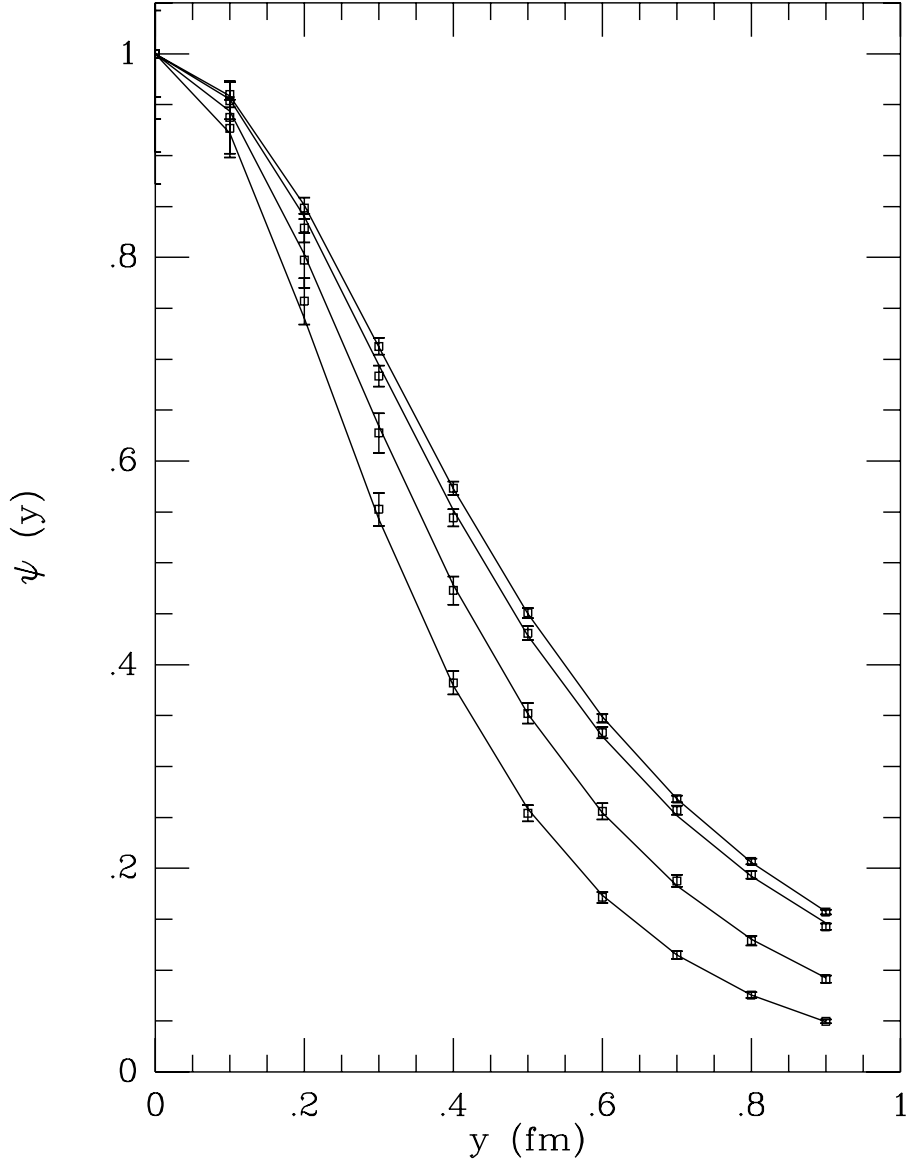

\begin{center}
\leavevmode
%\epsffile{pionbs.ps}
\end{center}
\caption{Pion Bethe Salpeter wave function measured at different
separations $\tau=0.5$, 0.75, 1.0 and 1.25 fm. All wave functions
are normalized to one at the origin. The lowest curve corresponds 
to $\tau=0.5$ fm. The solid curves show a parametrization used to 
extract the means square radii.}
\end{figure}
\vfill                  

\newpage
\begin{figure}
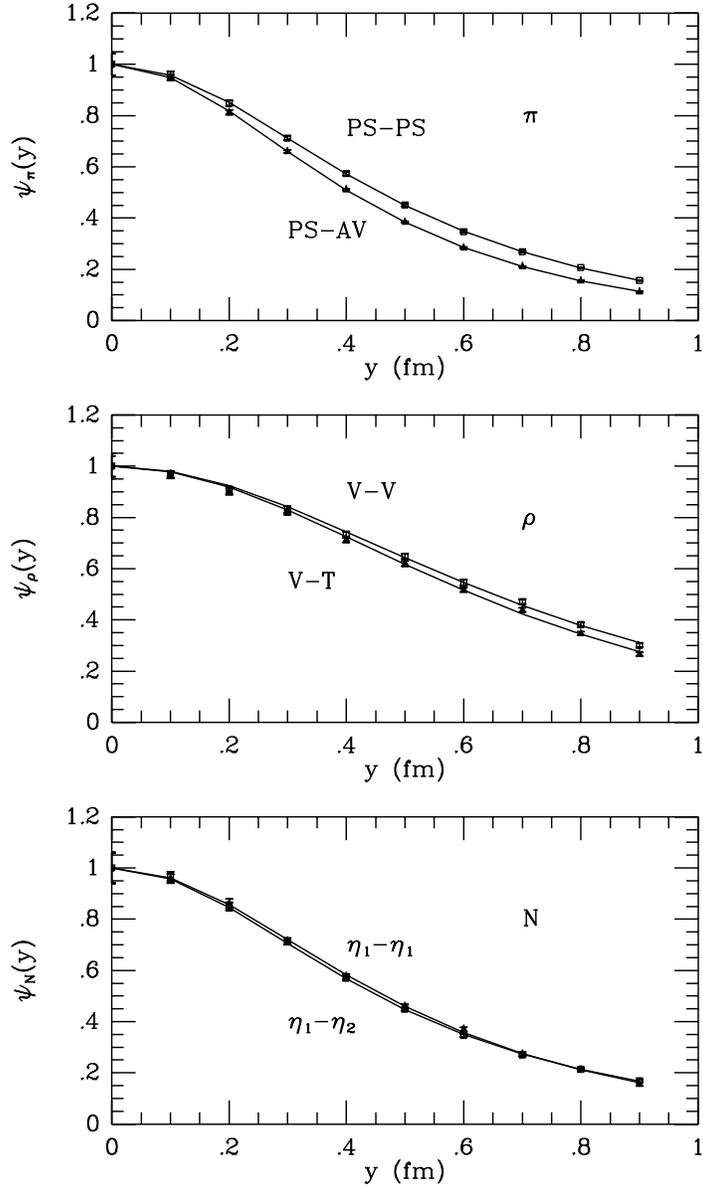

\begin{center}
\leavevmode
%\epsffile{offdiag.ps}
\end{center}
\caption{Bethe Salpeter wave functions for the pion, the rho meson
and the nucleon as determined from diagonal and off diagonal 
correlation functions at $\tau=1.25$ fm. The different correlators 
used are explained in the text. All wave functions are normalized to 
one at the origin.}
\end{figure}
\vfill

\newpage
\begin{figure}
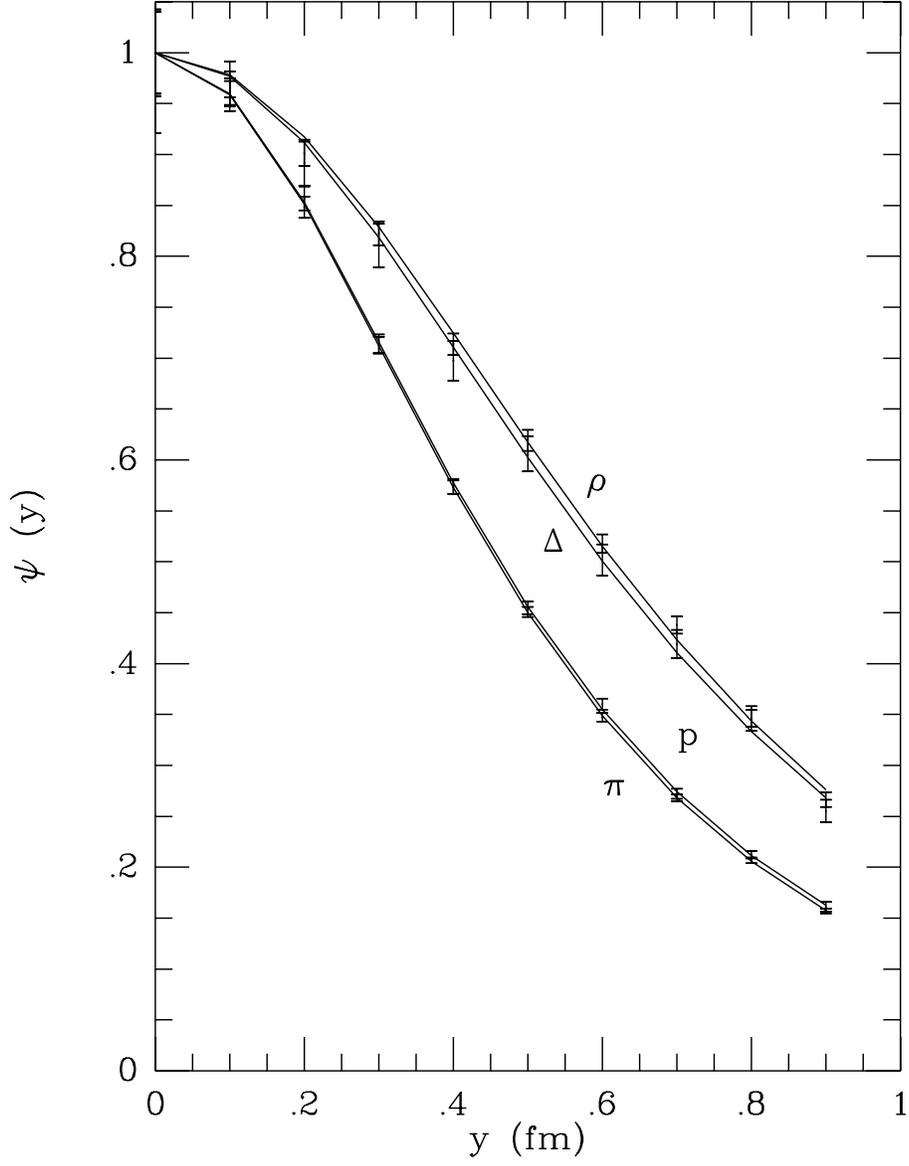

\begin{center}
\leavevmode
%\epsffile{hadbs.ps}
\end{center}
\caption{Bethe Salpeter amplitudes for the pion, the rho meson, the
nucleon and the delta measured in the random instanton model. The
wave functions are determined from diagonal correlation functions
at $\tau=1.25$ fm. They are normalized to one at the origin. The solid
curves show a parametrization of the data used to extract the 
corresponding means square radii.}
\end{figure}
\vfill                  
                                                        
\newpage
\begin{figure}
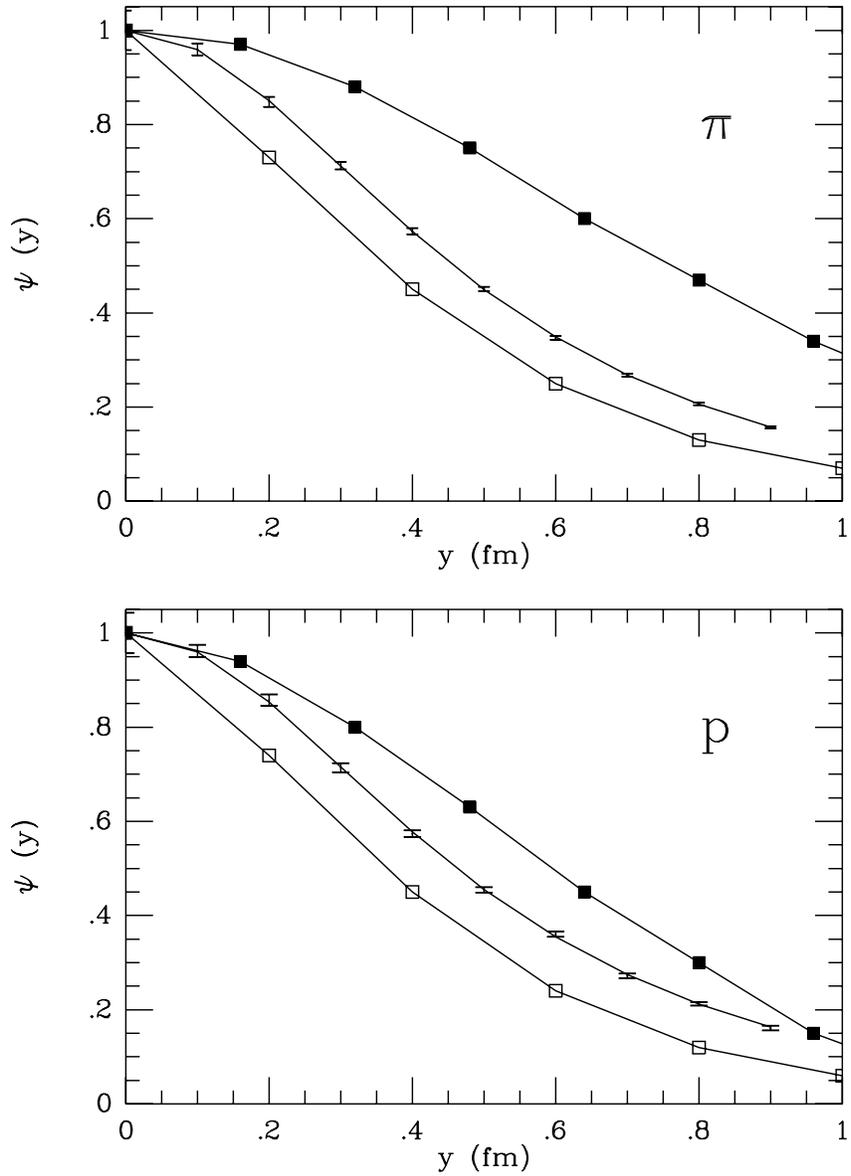

\begin{center}
\leavevmode
%\epsffile{compbs.ps}
\end{center}
\caption{Comparison of pion and proton Bethe Salpeter amplitudes
calculated in the random instanton model with the corresponding
results of lattice calculations reported in [2,3]. 
The lowest curve is the full lattice result, the curve in the middle 
shows the result in the instanton model and the upper curve corresponds 
to the lattice result after cooling.}
\end{figure}

\vfill                  
\end{document}